\documentclass{article}
\usepackage{amsmath,amssymb,graphicx}

\newcommand{\defeq}{\stackrel{\mathrm{def}}{=}}

\def\RR{\mathbb R}

\def\cC{\mathcal C}
\def\cD{\mathcal D}

\def\cG{\mathcal G}

\def\cP{\mathcal P}

\def\cV{\mathcal V}
\def\cX{\mathcal X}

\def\b1{\mathbf 1}

\def\bar{\overline}

\newcommand{\nop}[1]{}

\def\cY{{\mathcal Y}}
\def\cZ{{\mathcal Z}}

\newcommand{\qed}{\hfill$\square$\bigskip}
\newcommand{\raf}[1]{(\ref{#1})}
\newcommand{\proof}{\noindent {\bf Proof}.~~}

\newcommand{\hide}[1]{}

\newcommand{\VD}[1]{Dec$(#1,\cX)$}

\newtheorem{theorem}{Theorem}

\newtheorem{claim}{Claim}

\newtheorem{proposition}{Proposition}

\title{Characterization of the Vertices and Extreme Directions of the Negative Cycles Polyhedron and Hardness of Generating Vertices of $0/1$-Polyhedra}
\author{
Endre Boros\thanks {RUTCOR, Rutgers University, 640  Bartholomew
Road, Piscataway NJ  08854-8003; (boros@rutcor.rutgers.edu)}
\and
Khaled Elbassioni\thanks{Max-Planck-Institut f\"ur Informatik,
Saarbr\"ucken, Germany; (elbassio@mpi-sb.mpg.de)}
\and
Vladimir Gurvich\thanks {RUTCOR, Rutgers University, 640  Bartholomew
Road, Piscataway NJ  08854-8003; $\;\;\;\;$ (gurvich@rutcor.rutgers.edu)}
\and 
Hans Raj Tiwary\thanks{Universit\"at des Saarlandes, Saarbr\"ucken, D-66125 Germany; (hansraj@cs.uni-sb.de)} 
}

\begin{document}
\date{}
\maketitle
\begin{abstract}
Given a graph $G=(V,E)$ and a weight function on the edges $w:E\mapsto\RR$, we consider the polyhedron $P(G,w)$ of negative-weight flows on $G$, and get a complete characterization of the vertices and extreme directions of $P(G,w)$. As a corollary, we show that, unless $P=NP$, there is no output polynomial-time algorithm to generate all the vertices of a $0/1$-polyhedron. This strengthens the NP-hardness result of \cite{KBBEG06} for non 
$0/1$-polyhedra, and comes in contrast with the polynomiality of vertex enumeration for $0/1$-polytopes \cite{BL98}.

{\bf Keywords:}
Flow polytope, $0/1$-polyhedron, vertex, extreme direction, enumeration problem,
negative cycles, directed graph.
\end{abstract}

\section{Introduction}
A convex polyhedron $P\subseteq\RR^n$ is the the intersection of finitely many halfspaces, determined by the \emph{facets} of the polyhedron.
A \emph{vertex} or an \emph{extreme point} of $P$ is a point $v\in\RR^n$ which cannot be represented as a convex combination of two other points of $P$, i.e., there exists no $\lambda\in(0,1)$ and $v_1,v_2\in P$ such that $v=\lambda v_1+(1-\lambda)v_2$. A \emph{direction} of $P$ is a vector $d\in\RR^n$ such that $x_0+\mu d\in P$ whenever $x_0\in P$ and $\mu\ge 0$. An \emph{extreme direction} of $P$ is a direction $d$ that cannot be written as a conic combination of two other directions, i.e., 
there exist no non-negative numbers $\mu_1,\mu_2\in\RR_+$ and directions $d_1,d_2$ of $P$ such that $d=\mu_1 d_1+\mu_2 d_2$.   
Denote respectively by $\cV(P)$ and $\cD(P)$ the sets of extreme points and directions of polyhedron $P$. 
A bounded polyhedron, i.e., one for which $\cD(P)=\emptyset$ is called a \emph{polytope}.

\medskip

The well-known Minkowski-Weyl theorem states that any convex polyhedron can be represented as the Minkowski sum of the convex hull
of the set of its extreme points and the conic hull of the set of its extreme directions (see e.g. \cite{S86}). Furthermore, for pointed polyhedra, i.e., those that do not contain lines, this representation is unique. Given a polyhedron $P$ by its facets, obtaining the set $\cV(P)\cup\cD(P)$, required by the other representation, is a well-known problem, studied in the literature under different (but polynomially equivalent) forms, e.g. the \emph{vertex enumeration} problem\cite{BFM98}, the \emph{convex hull} problem \cite{ABS97} or the \emph{polytope-polyhedron problem} \cite{Lov92}. Clearly, the size of the extreme set $\cV(P)\cup\cD(P)$ can be (and typically is) exponential in $n$ and the number of facets $m$, and thus when we consider the computational complexity of the vertex enumeration problem, we are interested in \emph{output-sensitive} algorithms,
i.e., whose running time depends not only on $n,m$, but also on $|\cV(P)\cup\cD(P)|$. Alternatively, we may consider the following, polynomially equivalent, decision variant of the problem:

\begin{description}
\item [\VD{\cC(P)}:] Given a polyhedron $P$, represented by a system of linear inequalities, and
a subset $\cX\subseteq\cC(P)$, is $\cX=\cV(P)$?
\end{description} 
In this description, $\cC(P)$ could be either $\cV(P)$, $\cD(P)$, or $\cV(P)\cup\cV(P)$. 
It is well-known and also easy to see that the decision problems for $\cD(P)$ or for $\cV(P)\cup\cD(P)$ are equivalent to that for $\cV(P')$ where $P'$ is some polytope derived from $P$.
It is also well-known that if the decision problem is NP-hard, then no output polynomial-time algorithm can generate the elements of the set $\cC(P)$ unless P=NP (see e.g. \cite{BEGM07}). 

\medskip

The complexity of some interesting restrictions of these problems have already been settled. Most notably, it was shown in \cite{BL98}, that in the case of $0/1$-polytopes, i.e., for which $\cV(P)\subseteq\{0,1\}^n$, the problem of finding the vertices given the facets can be solved with polynomial delay (i.e. the time to produce each vertex is bounded by a polynomial in the input size) using a simple backtracking algorithm. 
Output polynomial-time algorithms also exist for enumerating the 
vertices of simple and simplicial polytopes \cite{AF92,AF96,BFM98}, network polyhedra and their duals \cite{P94}, and some other classes of polyhedra \cite{ADP03}.
More recently, it was shown in \cite{KBBEG06} that for general unbounded polyhedra problem \VD{\cV(P)} of generating the vertices of a polyhedron $P$ is NP-hard. On the other hand, for special classes of $0/1$-polyhedra, e.g. the polyhedron of $s$-$t$-cuts in general graphs \cite{GV95}, the polyhedra associated with the incidence matrix of bipartite graphs, and the polyhedra associated with $0/1$-network matrices \cite{BEGM07}, the  vertex enumeration problem can be solved in polynomial time using problem-specific techniques. This naturally raises the question whether there exists a general polynomial-time algorithm for the vertex enumeration of such polyhedra, extending the result of \cite{BL98} for $0/1$-polytopes. Here, we show that this is only possible if P=NP. Our result strengthens that in \cite{KBBEG06}, which did not apply to $0/1$-polyhedra, and uses almost the same construction, but goes through the characterization of the vertices of the polyhedron of negative weight-flows of a graph, defined in the next section. We show that this polyhedron could be highly unbounded, by also characterizing its extreme directions, and leave open the hardness of enumerating these directions, with its immediate consequences on the hardness of vertex enumeration for polytopes.  

\section{The polyhedron of negative-weight flows}\label{s1}
Given a directed graph  $G = (V,E)$
and a weight function  $w: E \rightarrow \RR$ on its arcs, consider the following polyhedron:  
\[
P(G,w)=\left\{y\in \RR^E~\left|~
\begin{array}{cl}\displaystyle (F)~~~\sum_{v:(u,v)\in
E}y_{uv}-\sum_{v:(v,u)\in E}y_{vu} ~=~ 0 &\forall~~ u\in V\\*[7mm]
\displaystyle (N)~~~~~~~~~~~~~~~~\sum_{(u,v)\in E} w_{uv}y_{uv} ~=~ 
-1&\\*[3mm]y_{uv}\geq 0 &\forall~~
(u,v)\in E\end{array}\right.\right\}.
\]

If we think of $w_{u,v}$ as the cost/profit paid for edge $(u,v)$ per unit of flow, then each point of $P(G,w)$ represents a \emph{negative-weight circulation} in $G$, i.e., assigns a non-negative flow on the arcs, obeying the \emph{conservation of flow} at each node of $G$, and such that total weight of the flow is strictly negative.

\begin{figure}[btp] 
\center\includegraphics[width=10cm]{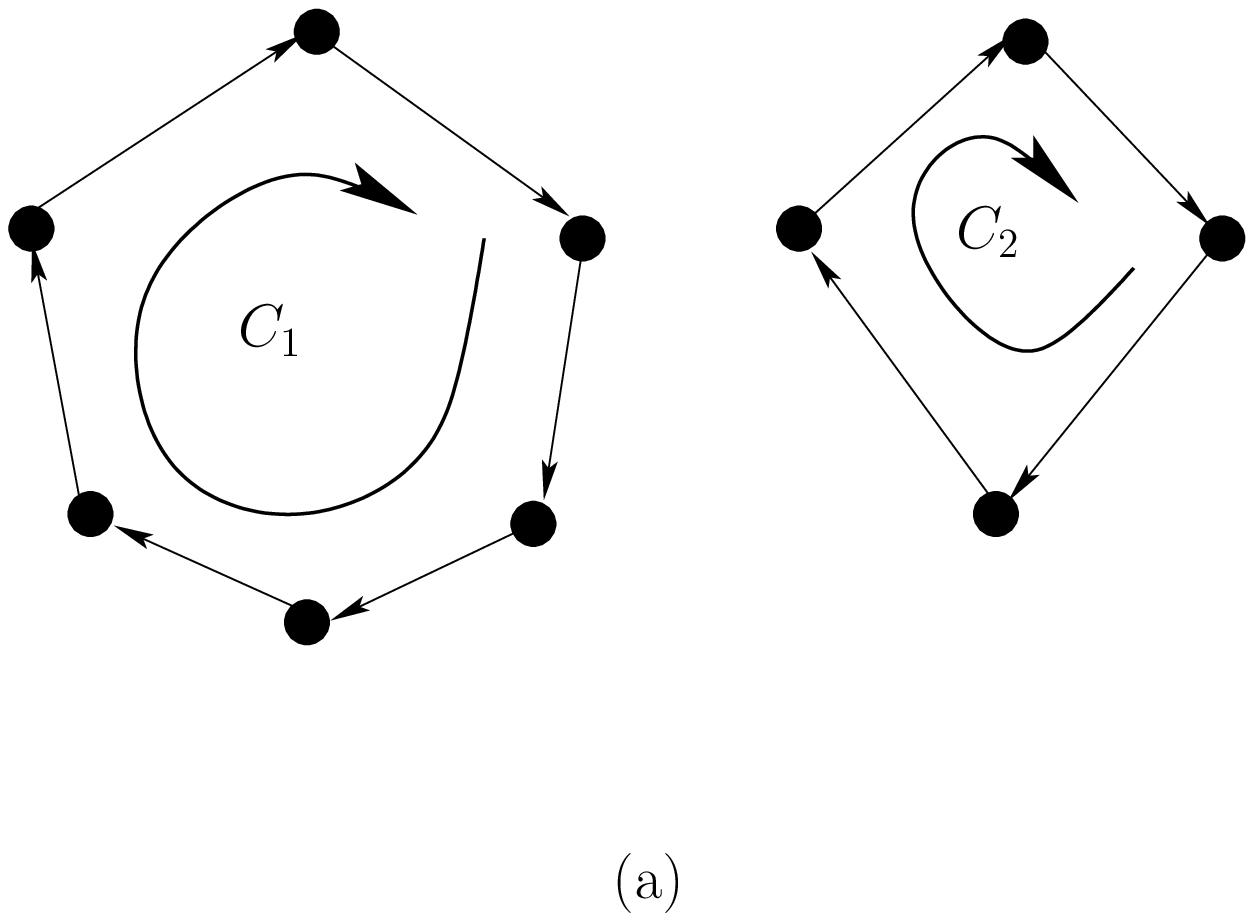}\\
\caption{$2$-cycle. }
\label{f1}
\end{figure}

A negative- (respectively, positive-, or zero-) weight cycle in $G$ 
is a directed cycle whose total weight is negative 
(respectively, positive, or zero). We represent a cycle $C$ by the subset of arcs appearing on the cycle, and denote by $V(C)$ the nodes of $G$ on the cycle (we assume all cycles considered to be directed and simple).
Let us denote the families of all 
negative, positive, and zero-weight cycles of $G$ 
by  $\cC^-(G,w),$ $\cC^+(G,w)$, and $\cC^0(G,w)$, respectively. 
Define a \emph{2-cycle} to be a pair of cycles $(C_1,C_2)$ 
such that $C_1\in\cC^-(G,w), C_2\in\cC^+(G,w)$ and $C_1\cup C_2$ does not contain any other cycle of $G$. It is not difficult to see that a $2$-cycle is either the edge-disjoint union of a negative cycle $C_1$ and a positive cycle $C_2$, or the edge-disjoint union of 3 paths $P_1,$ $P_2$ and $P_3$ such that $C_1=P_1\cup P_2$ is a negative cycle, and $C_2=P_1\cup P_3$ is a positive cycle (see Figure \ref{f1}).  
In the next section, we show that the set of vertices $\cV(P(G,w))$ are in one-to-one correspondence with 
the set of negative cycles $\cC^-(G,w),$ while the set of extreme directions $\cD(P(G,w))$ is in one-to-one correspondence with the set
$\cC^0(G,w)\cup\{(C,C')~:~(C,C')\mbox{ is a $2$-cycle})\}.$ 

\section{Characterization of vertices and extreme directions of $P(G,w)$}
For a subset $X\subseteq E$, and a weight function $w:E\mapsto\RR$, we denote by $w(X)=\sum_{e\in X}w_e$, the total weight of $X$. For $X\subseteq E$, we denote by $\chi(X)\in\{0,1\}^E$ the characteristic vector of $X$: $\chi_e(X)=1$ if and only if $e\in X$, for $e\in E$. 

\begin{theorem}\label{t1}
Let $G=(V,E)$ be a directed graph and $w: E \rightarrow \RR$ be a 
real weight on the arcs. Then
\begin{eqnarray}
\label{e1}
\cV(P(G,w))&=&\left\{\frac{-1}{w(C)}\chi(C):~C\in\cC^-(G,w)\right\},\\
\label{e2}
\cD(P(G,w))&=&\cD_1\cup\cD_2,
\end{eqnarray}
where 
\begin{eqnarray*}
\cD_1&=&\{\frac{1}{|C|}\chi(C):~C\in\cC^0(G,w)\},\\
\cD_2&=&\{\mu_{C_1,C_2}\chi(C_1)+\mu'_{C_1,C_2}\chi(C_2)~:~(C_1,C_2)\mbox{ is a 2-cycle}\},
\end{eqnarray*}
and 
$$
\mu_{C_1,C_2}=\frac{w(C_2)}{w(C_2)|C_1|-w(C_1)|C_2|},~\mu'_{C_1,C_2}=\frac{-w(C_1)}{w(C_2)|C_1|-w(C_1)|C_2|}.
$$
are non-negative numbers computed from cycles $C_1$ and $C_2.$
\end{theorem}
\proof
Let $m=|E|$ and $n=|V|$. We first prove \raf{e1}. It is easy to verify that any element $y\in\RR^E$ of the set on the right-hand side of \raf{e1} belongs to $P(G,w)$. Moreover, any such $x=-\chi(C)/w(C)$, for a cycle $C$, is a vertex of $P(G,w)$ since there are $m$ linearly independent inequalities of $P(G,w)$ tight at $x$, namely: the conservation of flow equations at $|C|-1$ vertices of $C$, the equation $\sum_{e\in C}w_e y_e=-1$, and $m-|C|$ equations $y_e=0$, for $e\in E\setminus C$. 

To prove the opposite direction, let $y\in \RR^E$ be a vertex of $P(G,w)$. Let $Y=\{e\in E:~y_e>0\}$. The proof follows from the following 3 claims.

\begin{claim}\label{cl1}
The graph $(V,Y)$ is the disjoint union of strongly connected components.
\end{claim}
\proof
Consider an arbitrary strongly connected component $X$ in this graph, and let $X^-$ be the set of components reachable from $X$ (including $X$). Summing the conservation of flow equations corresponding to all the nodes in $X^-$ implies that all arcs going out of $X^-$ have a flow of zero. 
\qed

\medskip

\begin{claim}\label{cl2} 
There exists no cycle $C\in\cC^0(G,w)$ such that $C\subseteq Y$.
\end{claim}  
\proof
If such a $C$ exists, we define two points $y'$ and $y''$ as follows. 
$$
y_e'=\left\{
\begin{array}{ll}
y_e+\epsilon,& \mbox{if $e\in C$}\\
y_e, &\mbox{otherwise,}
\end{array}
\right.
\ \ \ \ 
y_e''=\left\{
\begin{array}{ll}
y_e-\epsilon,& \mbox{if $e\in C$}\\
y_e, &\mbox{otherwise,}
\end{array}
\right.
$$
for some sufficiently small $\epsilon>0$. Then $y',y''$ clearly satisfy (F). Moreover, (N) is satisfied with $y'$ since 
$$\sum_{e\in E}w_e y'_e=\sum_{e\not\in C}w_ey_e+\sum_{e\in C}w_e(y_e+\epsilon)=\sum_{e\in E}w_ey_e+w(C)\epsilon=-1.$$
Similarly for $y''.$ Thus $y',y''\in P(G,w)$ and $y=(y'+y'')/2$ contradicting that $y$ is a vertex.
\qed

\medskip

\begin{claim}\label{cl3} 
There exist no distinct cycles $C_1,C_2\in\cC^-(G,w)\cup\cC^+(G,w)$ such that $C_1\cup C_2\subseteq Y$.
\end{claim}  
\proof
If such $C_1$ and $C_2$ exist, we define two points $y'$ and $y''$ as follows. 
$$
y_e'=\left\{
\begin{array}{ll}
y_e+\epsilon_1,& \mbox{if $e\in C_1\setminus C_2$}\\
y_e+\epsilon_2,& \mbox{if $e\in C_2\setminus C_1$}\\
y_e+\epsilon_1+\epsilon_2,& \mbox{if $e\in C_1\cap C_2$}\\
y_e, &\mbox{otherwise,}
\end{array}
\right.
\ \ \ \ 
y_e'=\left\{
\begin{array}{ll}
y_e-\epsilon_1,& \mbox{if $e\in C_1\setminus C_2$}\\
y_e-\epsilon_2,& \mbox{if $e\in C_2\setminus C_1$}\\
y_e-\epsilon_1-\epsilon_2,& \mbox{if $e\in C_1\cap C_2$}\\
y_e, &\mbox{otherwise,}
\end{array}
\right.
$$
where $\epsilon_1=-\frac{w(C_2)}{w(C_1)}\epsilon_2$, for some sufficiently small $\epsilon_2>0$ (in particular, to insure non-negativity of $y',y''$, $\epsilon_2$ must be upper bounded by the minimum of $\min\{y_e: e\in C_2\setminus C_1\}$, $\frac{|w(C_1)|}{|w(C_2)|}\min\{y_e: e\in C_1\setminus C_2\}$, and $\frac{|w(C_1)|}{|w(C_1)-w(C_2)|}\min\{y_e: e\in C_1\cap C_2\}$). Then it is easy to verify that $y',y''$ satisfy (F). Moreover, (N) is satisfied with $y'$since 
\begin{eqnarray*}
\sum_{e\in E}w_e y'_e&=&\sum_{e\not\in C_1\cup C_2}w_ey_e+\sum_{e\in C_1\setminus C_2}w_e(y_e+\epsilon_1)+\sum_{e\in C_2\setminus C_1}w_e(y_e+\epsilon_2)\\
&&+\sum_{e\in C_1\cap C_2}w_e(y_e+\epsilon_1+\epsilon_2)=\sum_{e\in E}w_ey_e+w(C_1)\epsilon_1+w(C_2)\epsilon_2=-1.\end{eqnarray*}
Similarly for $y''.$ Thus $y',y''\in P(G,w)$ and $y=(y'+y'')/2$ contradicting that $y$ is a vertex of $P(G,w)$.
\qed

\medskip

The above 3 claims imply that the graph $(V,Y)$ consists of a single cycle $C$ and a set of isolated vertices $V\setminus V(C)$. Thus $y_e=0$ for $e\not\in C$. By (F) we get that $y_e$ is the same for all $e\in C$, and by (N) we get that $y_e=-1/w(C)$ for all $e\in C$, and in particular that $C\in\cC^-(G,w)$. This completes the proof of \raf{e1}.    
\medskip

We next prove \raf{e2}. As is well-known, the extreme directions of $P(G,w)$ are in one-to-one correspondence with the vertices of the polytope $P'(G,w)$, obtained from 
$P(G,w)$ by setting the right-hand side of (N) to $0$ and adding the normalization constraint $(N')~:~\sum_{e\in E}y_e=1$.

We first note as before that every element of $\cD_1\cup\cD_2$ is a vertex of $P'(G,w)$. Indeed, if $y\in\cD_2$ is defined by a $2$-cycle $(C_1,C_2)$, then there are $m$ linearly independent inequalities tight at $y$. To see this, we consider two cases: (i) When $C_1$ and $C_2$ are edge-disjoint, then there are $|C_1|-1$ and $|C_2|-1$ equations of type (F), normalization equations $(N)$ and $(N')$, and $m-|C_1|-|C_2|$ non-negativity inequalities for $e\in E\setminus (C_1\cup C_2)$. (ii) Otherwise, $C_1\cup C_2$ consists of 3 disjoint paths $P_1,P_2,P_3$ of, say $m_1,m_2$ and $m_3$ arcs, respectively. Then $C_1\cup C_2$ has $m_1+m_2+m_3-1$ giving $m_1+m_2+m_3-2$ linearly independent equation of type (F), which together with $(N)$, $(N')$ and $m-m_1-m_2-m_3$ non-negativity constraints for $e\in E\setminus (C_1\cup C_2)$ uniquely define $y$.

Consider now a vertex $y$ of $P'(G,w)$. Let $Y=\{e\in E:~y_e>0\}$. Clearly, Claim \ref{cl1} is still valid for $Y$. On the other hand, Claims \ref{cl2} and \ref{cl3} can be replaced by the following two claims.

\begin{claim}\label{cl4} 
There exist no 3 distinct cycles $C_1,C_2,C_3$ such that $C_1\in\cC^-(G,w)$, $C_2\in\cC^+(G,w)$, and $C_1\cup C_2\cup C_3\subseteq Y$.
\end{claim}  
\proof
If such $C_1,$ $C_2$ and $C_3$ exist, we define two points $y'$ and $y''$ as follows: $y'_e=y_e+\sum_{i=1}^3\epsilon_i\chi_e(C_i)$ and $y''_e=y_e-\sum_{i=1}^3\epsilon_i\chi_e(C_i)$, for $e\in E$, where $\epsilon_3>0$ is sufficiently small, and $\epsilon_1$ and $\epsilon_2$ satisfy
\begin{eqnarray}
\epsilon_1 w(C_1)+\epsilon_2 w(C_2)&=&-\epsilon_3 w(C_3)\nonumber\\
\epsilon_1 |C_1|+\epsilon_2 |C_2|&=&-\epsilon_3|C_3|.
\label{e4}
\end{eqnarray}
Note that $\epsilon_1$ and $\epsilon_2$ exist since $\alpha\defeq w(C_1)|C_2|-w(C_2)|C_1|<0$. Furthermore, since $\epsilon_1=(w(C_2)|C_3|-w(C_3)|C_2|)\epsilon_3/\alpha$ and $\epsilon_2=(w(C_3)|C_1|-w(C_1)|C_3|)\epsilon_3/\alpha$, we can select $\epsilon_3$ such that $y',y''\ge 0$.
By definition of $y'$ and $y'',$ they both satisfy (F), and by \raf{e4} they also satisfy $(N)$ and $(N')$. However, $(y'+y'')/2=y$ contradicts that $y\in\cV(P'(G,w))$.
\qed 

\begin{claim}\label{cl5} 
There exist no 2 distinct cycles $C_1,C_2$ such that $C_1,C_2\in\cC^0(G,w)$, and $C_1\cup C_2\subseteq Y$.
\end{claim}  
\proof
If such $C_1$ and $C_2$ exist, we define two points $y'$ and $y''$ as follows: $y'_e=y_e+\epsilon_1\chi_e(C_1)+\epsilon_2\chi_e(C_2)$ and 
$y''_e=y_e-\epsilon_1\chi_e(C_1)-\epsilon_2\chi_e(C_2)$, for $e\in E$, where $\epsilon_2>0$ is sufficiently small, and $\epsilon_1=-\epsilon_2|C_2|/|C_1|$. Then $y',y''\in P'(G,w)$ and $y=(y'+y'')/2$.
\qed

\medskip

As is well-known, we can decompose $y$ into the sum of positive flows on cycles, i.e., write $y=\sum_{C\in\cC'}\lambda_C \chi(C)$, where $\cC'\subseteq\cC^-(G,w)\cup\cC^+(G,w)\cup\cC^0(G,w)$, and $\lambda_C>0$ for $c\in\cC'$. It follows from Claim \ref{cl4} that $|\cC'|\le 2$. 
Using $(N)$, we get $\sum_{c\in\cC'}\lambda_C w(C)=0$, which implies by Claim \ref{cl5} that either $\cC'=\{C\}$ and $w(C)=0$ or $\cC'=\{C_1,C_2\}$ and
$w(C_1)<0$, $w(C_1)>0$. In the former case, we get that $y\in\cD_1$, and in the latter case, we get by Claim \ref{cl4} that $(C_1,C_2)$ is a $2$-cycle, and hence, that $y\in\cD_2$.
\qed

In the next section we construct a weighted directed graph $(G,w)$ in which all negative cycles have unit weight. We show that
generating all negative cycles of $G$ is NP-hard, thus implying by Theorem \ref{t1} 
that generating all vertices of $P(G,w)$ is also hard. 

\section{Generating all vertices of a $0/1$-polyhedron is hard}

Let us now show that the following problem is CoNP-complete:
\begin{description}
\item[VE-$0/1$:] Given a polyhedron $P=\{x\in\RR^n|~Ax\le b\}$, where $A\in\RR^{m\times n}$, $b\in\RR^m$, and $\cV(P)\subseteq\{0,1\}^n$, and a subset $\cX\subseteq\cV(P)$, decide if $\cX=\cV(P).$ 
\end{description}
 
Then, no algorithm can generate all elements of $\cV(P)$
in incremental or total polynomial time, unless P$=$NP. 

\smallskip

\begin{theorem}\label{t2}
Problem VE-$0/1$ is NP-hard.
\end{theorem}
\proof
The construction is essentially the same as in \cite{KBBEG06}; only the weights change. We include a sketch here.

We reduce the problem from
the CNF satisfiability problem: Is there a
truth assignment of $N$ binary variables satisfying all clauses of a given
conjunctive normal form $\phi(x_1,\ldots,x_N) = C_1\wedge \ldots \wedge C_m,$
where each $C_j$ is a disjunction of some literals in $\{x_1,\bar
x_1,\ldots,x_n,\bar x_n\}$?

Given a CNF $\phi$, we construct a weighted directed graph $G=(V,E)$ on $|V|=5\sum_{j=1}^m|C_j|+m-n+1$ vertices and
$|E|=6\sum_{j=1}^m|C_j|+1$ arcs (where $|C_j|$ denotes the number of
literals appearing in clause $C_j$) as follows.
For each literal $\ell=\ell^j$ appearing in clause $C_j$, we
introduce two paths of three arcs each: $\cP(\ell)=(p(\ell),a(\ell),b(\ell),q(\ell))$, and $\cP'(\ell)=(r(\ell),b'(\ell),a'(\ell),s(\ell))$. The weights of these arcs are set as follows: 
$$
\begin{array}{lll}
w((p(\ell),a(\ell)))=\frac{1}{2}, & w((a(\ell),b(\ell)))=-\frac{1}{2}, & w((b(\ell),q(\ell)))=0,\\\\*[2mm]
w((r(\ell),b'(\ell)))=0, &w((b'(\ell),a'(\ell)))=-\frac{1}{2},& w((b'(\ell),s(\ell)))=\frac{1}{2}.
\end{array}
$$
           
\begin{figure}[btp] 
\center\includegraphics[width=12cm]{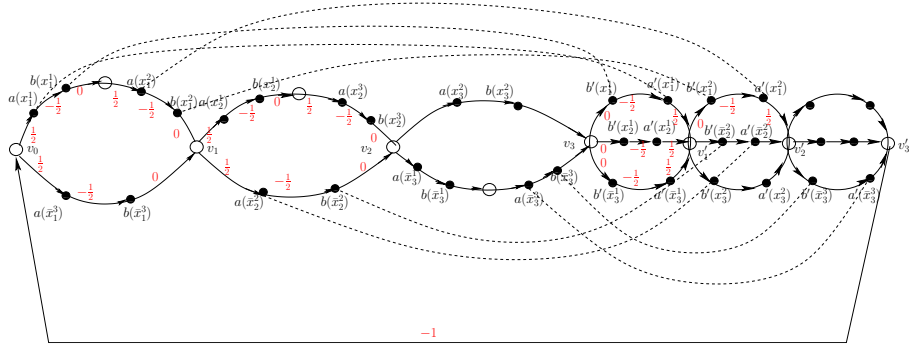}\\
\caption{An example of the graph construction in the proof of Theorem \ref{t2} with CNF $C=(x_1\vee
x_2\vee\overline{x}_3)\wedge(x_1\vee\overline{x}_2\vee x_3)\wedge(\overline{x}_1\vee x_2\vee\overline{x}_3)$. }
\label{f2}
\end{figure}

These edges are connected in $G$ as follows (see Figure \ref{f2} for an example):
$$G= v_0~\cG_1~v_1 ~ \cG_2 ~v_2 \ldots ~v_{n-1}~ \cG_n~v_n~\cG_1'~v_1'~\cG_2'~v_2'\ldots~v_{m-1}'~\cG_m'~v_m',$$
where $v_0,v_1,\ldots,v_{n},v_1',\ldots,v_{m-1}',v_m'  $ are
distinct vertices, each $\cG_i=\cY_i\vee\cZ_i$, for $i=1,\ldots,n$,
consists of two parallel chains $\cY_i=\wedge_{j}\cP(x_i^j)$
and $\cZ_i=\wedge_{j}\cP(\bar x_i^j)$ between $v_{i-1}$ and $v_i$,
and each $\cG_j'=\vee_{i=1}^{|C_j|}\cP'(\ell_i^j)$, for $j=1,\ldots,m$, where $\ell^j_1,\ell^j_2,\ldots$ are the literals appearing in $C_j$.

Finally we add the arc $(v_m',v_0)$ with weight $-1$, and \emph{identify} the pairs of nodes $\{a(\ell),a'(\ell)\}$ and $\{b(\ell),b'(\ell)\}$ for all $\ell$, (i.e. $a(\ell)=a(\ell')$ and $b(\ell)=b(\ell')$ define the same nodes).

Let $P(G,w)$ be the polyhedron defined by the graph $G$ and the arc weights $w$. We shall argue now that all negative cycles of $G$ have weight -1. This implies by Theorem \ref{t1} that all vertices of $P(G,w)$ are $0/1$. 

Clearly the arcs $(a(\ell),b(\ell))$ and $(b'(\ell),a'(\ell))$ form a directed cycle of total weight $-1$, for every literal occurrence $\ell$. 
There are $\sum_{j=1}^m|C_j|$ such cycles, corresponding to a subset $\cX\subseteq\cV(P(G,w))$. 

Call a cycle of $G$ long if it contains the vertices $v_0,v_1,\ldots,v_{n},v_1',\ldots,v_{m-1}',v_m'$. Any long cycle has weight $-1$. The crucial observation is the following.
 
\begin{claim}\label{cl6}
Any negative cycle $C\in\cC^-(G,w)\setminus\cX$ must be long.
\end{claim}
\proof
Consider any cycle $C\not\in\cX$, and let us write the traces of the nodes visited on the cycle (dropping the literals, and considering $a,a'$ and $b,b'$ as different copies), without loss of generality as follows:
$$
p~a~b~p~a~b~p~\cdots~a~a'~s~b'~a'~s~b'~\cdots~b'~b~p~a~b~\cdots~p. 
$$
Note that the sequences $a'~a$ and $b~b'$ are not allowed since otherwise $C$ contains a cycle from $\cX$. 

Let us compute the distance (i.e., the total weight) of each node on this sequence starting from the initial $p$. Call the subsequences $a'~a'$ and $b~b'$, $a$- and $b$-jumps respectively. Then it is easy to verify that each $a$-jump causes the distance to eventually increase by $1$ while each $b$-jump keeps the distance at its value. More precisely, the distance at a node $x$ in the sequence is given by $d(x)=t+d_0-\delta(x)$, where $t$ is the number of $a$-jumps appearing upto $x$, and
$$
d_0=\left\{
\begin{array}{ll}
0&\mbox{if $x\in\{p,s\}$},\\
\frac{1}{2}&\mbox{if $x=a$},\\
-\frac{1}{2}&\mbox{if $x=a'$},\\
0&\mbox{if $x=b=b'$},\\
\end{array}
\right.
~~\delta(x)=\left\{
\begin{array}{ll}
1& \mbox{if arc $(v_m',v_0)$ appears on the path from $p$ to $x$}\\
0& \mbox{otherwise}. 
\end{array}
\right.
$$
One also observes that, if the sequence has a $b$-jump, then it must also contain an $a$-jump. Thus it follows from the definition of $d(x)$ that any cycle with a jump must be non-negative. So the only possible negative cycle not in $\cX$ must be long. 
\qed

By Claim \ref{cl6}, checking of $\cV(P(G,w))=\cX$ is equivalent to checking if $G$ has a long cycle. It is easy to see that the latter condition is equivalent to the non-satisfiability of the input CNF formula $\phi$ (see \cite{KBBEG06} for more details).  
\qed

\medskip

Thus, it is NP-hard to generate all vertices of a $0/1$-polyhedron.
However, \raf{e2} shows that 
the above construction cannot be used to 
imply the same hardness result for polytopes, since 
the numbers of positive and negative cycles 
can be exponential and, hence, polyhedron 
$P(G,w)$ can be highly unbounded. In fact, for the negative cycle polyhedron arising in the construction of Theorem \ref{t2}, we have the following. 

\begin{proposition}\label{p1}
For the directed graph $G=(V,E)$ and weight $w: E \rightarrow \RR$ used in the proof of Theorem \ref{t2}, both sets $\cD(P(G,w))$ and $\cV(P(G,w))\cup\cD(P(G,w))$ can be generated in incremental polynomial time. 
\end{proposition}
\proof
This follows from the fact that for every positive cycle in $G$ there is a negative cycle, edge-disjoint from it, and vice versa (assuming no clause consists of only one literal) as one can easily verify. Hence, the number of 2-cycles and thus the number of extreme directions of $P(G,w)$ satisfy $|\cD(P(G,w))|\ge\max\{|\cC^+(G,w)|,$ $|\cC^-(G,w)|\}+|\cC^0(G,w)|$. Thus $\cD(P(G,w)$ and $\cV(P(G,w))\cup\cD(P(G,w))$  can be generated by generating all cycles of $G$, which can be done with polynomial delay \cite{RT75}.
\qed

\begin{figure}[btp] 
\center\includegraphics[width=8cm]{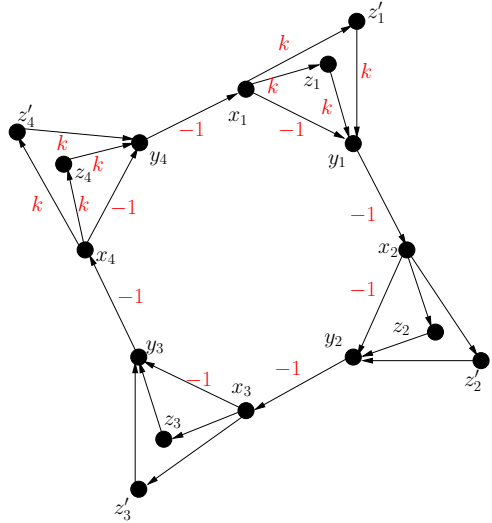}\\
\caption{An example where there are exponentially more positive cycles than 2-cycles ($k=4$). }
\label{f3}
\end{figure}

However, it is open whether the same holds for general graphs. In fact, there exist weighted graphs in which the number of positive cycles is exponentially larger than the number of 2-cycles. Consider for instance, a graph $G$ composed of a directed cycle $(x_1,y_1,\ldots,x_k,y_k)$ of length $2k$, all arcs with weight $-1$, and $2k$
additional paths $\cP_1,\cP_1',\ldots,\cP_k,\cP_k'$ where $\cP_i=(x_i,z_i,y_i)$ and $\cP_i'=(x_i,z'_i,y_i)$, of two arcs each going the same direction parallel
with every second arc along the cycle, each having a weight of $2k$ (see Figure \ref{f3} for an example with $k=4$). Then we
have more than $2^k$ positive cycles, but only $2k$ 2-cycles. Note that proving that enumerating 2-cycles of a given weighted graph is NP-hard, will imply the same for the vertex enumeration problem for polytopes, whose complexity remains open.

\bibliographystyle{amsplain}
\bibliography{vertexgen}
  
\end{document}